\documentclass{KapProc}
\usepackage{graphicx}
\kluwerbib
\begin{document}
\articletitle{QSO host galaxy star formation history from multicolour data} 
\author{
 Knud Jahnke$^1$, Bj\"orn Kuhlbrodt$^2$, Eva \"Orndahl$^3$, Lutz Wisotzki$^4$}
\affil{$^{1,2}$Hamburger Sternwarte; $^3$Astronomiska Observatoriet Uppsala;
  $^4$Universit\"at Potsdam}
\email{$^{1}$kjahnke@uni-hamburg.de,$^{2}$bkuhlbrodt@uni-hamburg.de,$^{3}$eva@astro.uu.se,\\
$^{4}$lutz@astro.physik.uni-potsdam.de,
}
\def\farcs{\hbox{$.\!\!^{\prime\prime}$}}
\begin{abstract}
We investigate multicolour imaging data of a complete sample of low
redshift ($z<0.2$) QSO host galaxies. The sample was imaged in four
optical (\textsl{BVRi}) and three near-infrared bands (\textsl{JHKs}),
and in addition spectroscopic data is available for a majority of the
objects.

We extract host luminosities for all bands by means of two-di\-men\-sio\-nal
modeling of galaxy and nucleus. Optical and optical-to-NIR colours
agree well with the average colours of inactive early type
galaxies. The six independent colours are used to fit population
synthesis models. We assess the presence of young populations in the
hosts for which evidence shows to be very weak.
\end{abstract}

\section{Goals}
For an assessment of galaxy-formation timescales QSO hosts play a
vital role, due to their obvious connection to black hole
formation. Dating the nuclear activity and possibly connecting this to
external events in the galaxy can help to decide on merger scenarios
and the triggering mechanism for activity.

With this work we wanted to start an assessment of the stellar content
of host galaxies. By decomposing the host into stellar components we
will be able in the future to make detailed comparisons to inactive
galaxies.

\section{Why multicolour data?}

In QSO host galaxy studies using single band or single optical--near
infrared (NIR)
colours is sufficient to characterise morphological properties like
galaxy types, host and nuclear luminosities, apparent signs of
interaction or to conduct environment studies (e.g.\ McLeod \& Rieke
1995, Percival et al. 2001). Optical colours or spectra permit the
characterisation of the dominant stellar population or allows
assessment of black hole masses (McLure et al. 1999, Boisson et
al. 2000). The NIR on the other hand yields, for low $z$, the best
contrast of host against nucleus and allows to assess the
mass-to-light ratio of a host.

For the separation of an SED into stellar populations of different
ages, using only optical information becomes insufficient for a unique
solution. In the NIR the emission of young populations rapidly
decreases and old populations dominate. Thus for a study of the
stellar components information about the entire SED from the optical
to NIR wavelength range is needed.

For luminous AGN the spectral separation of nuclear and host
components is very difficult and at the moment largely dependent on
subjective or ad hoc criteria for the nuclear component. The S/N
requirements limit studies to small redshifts and low nuclear
luminosities, as the acquisition of spectra becomes very expensive,
requiring 8m-class telescopes already at $z=0.2$. While the quality of
the spectral separation methods might change in the future the
now available two-dimensional modeling software for QSO hosts allows a
detailed and solid assessment of host galaxy fluxes and thus colour
information with a high degree of reliability.

\section{Sample \& observations}

We have compiled a sample of 20 objects with $z<0.2$, drawn from the
Hamburg/ESO survey (HES) for luminous QSOs (Wisotzki et al. 2000). The
HES is a flux limited objective-prism survey, with a limiting nuclear
magnitude \textsl{B}$_\mathrm{lim} \sim 17.5$ depending on the field,
designed to detect QSOs solely on basis of their spectral
properties. Thus unlike samples from many other QSO surveys, the
sample is not biased against extended objects.

\begin{figure}[bt]
\begin{minipage}[t][4.5cm][t]{7cm}
\includegraphics[bb = 52 563 366 771,clip,angle=0,height=4.5cm]{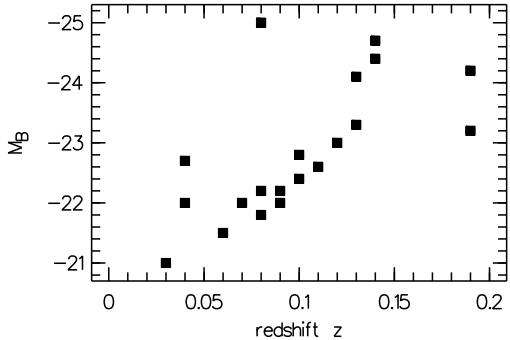}
\end{minipage}
\hfill
\begin{minipage}[b][4.5cm][t]{4.5cm}
\caption{\label{z_b} Sample properties. Absolute \textsl{B} band
magnitude of QSOs as a function of redshift.}
\end{minipage}
\end{figure}

The sample used is a complete sample from a sky area of
611$\;$deg$^2$, a low-$z$ subsample of a sample defined by K\"ohler et
al. (1997) to study the luminosity function of QSOs.
Distribution in redshift and absolute magnitude are
shown in Fig.\ \ref{z_b}, the sample represents moderately luminous
QSOs when compared to the total population at all redshifts. The
radio properties of most of the objects in the sample are not yet
known, but as a subset of the QSO population most will be radio-quiet.

For all 20 objects we have acquired \textsl{BVRiJHKs} broadband
photometry to evenly sample the SED over the optical--NIR wavelength
interval. With three NIR bands we get some redundancy in the NIR to
stabilise the stellar population fits. In addition we can make
comparisons of sample properties to samples at higher redshift without
the need for K-corrections. The \textsl{B} band images were integrated
30$\,$s at the ESO 3.6m telescope (EFOSC2), \textsl{VRi} images
300--1200$\,$s at ESO Danish 1.5m (DFOSC), and \textsl{JHKs}
160--900$\,$s at ESO NTT (SOFI). In addition we have available optical
spectra (3800--7500$\;$\AA) for 14 of the objects, taken with the ESO
3.6m telescope.

\section{Fitting stellar populations}

\begin{figure}[tb]
\includegraphics[bb = 75 565 366 771,clip,angle=0,height=4.2cm]{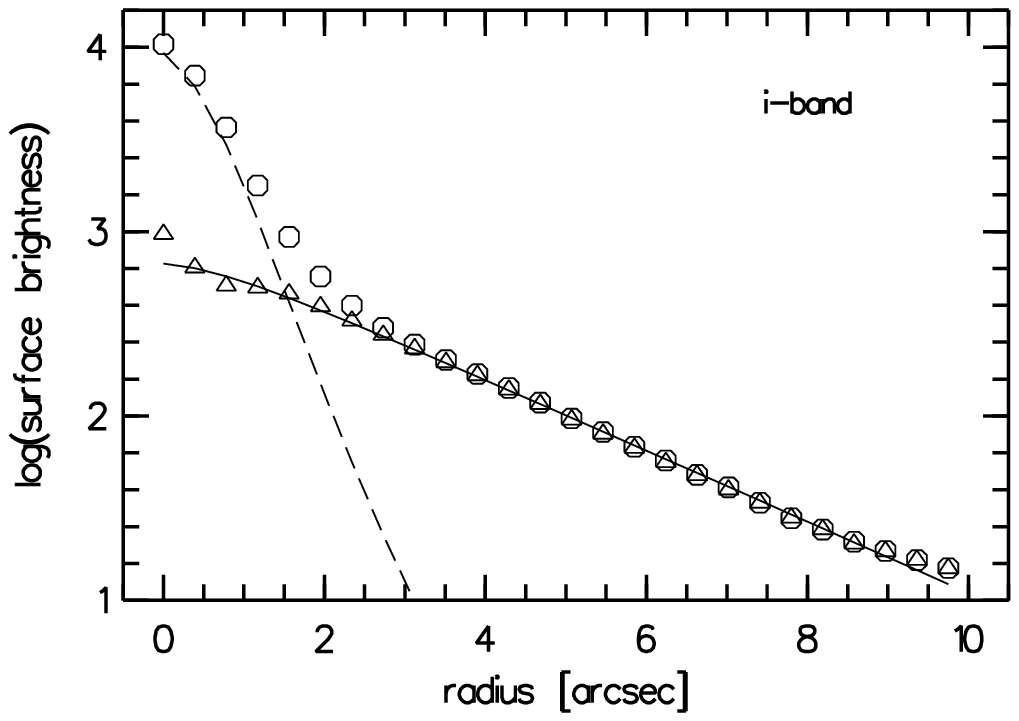}  
\includegraphics[bb = 75 565 366 771,clip,angle=0,height=4.2cm]{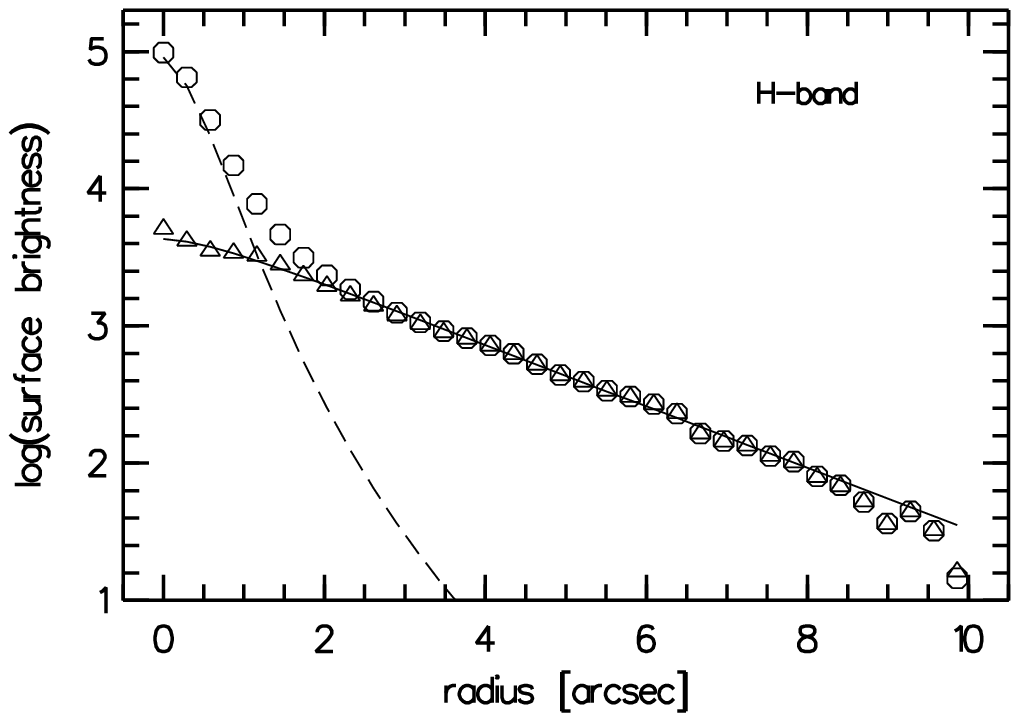}  
\caption{\label{sepa}
Radial profiles of HE$\,$1310--1051 in \textsl{i} band (left) and
\textsl{H} band (right). Circles: data points, dashed line: best
fitting nucleus, solid line: best fitting disk model, triangles:
residual galaxy light.}
\end{figure}

The nuclear contribution of the total QSO light has to be separated
from the stellar light. We have developed a package for simultaneous
two-dimensional modelling of a parametrised host model and the nuclear
contribution. Luminosities for the hosts are determined from radial
flux growth curves of the images, after subtracting the nuclear model
resulting from the best fit (Fig.\ \ref{sepa}). More details about the
modeling are given in the contribution by B.\ Kuhlbrodt et al.\
(these proceedings).

We could produce colours for 18 of the 20 objects. In the two
remaining cases the separation was not yielding unique solutions due
to highly disturbed morphological structure. We excluded these two
from further analysis.

For the optical spectra we are currently developing a two-dimensional
separation method similar to the imaging case. For some objects the
current program already yields host spectra almost free of broad
emission line components from the nucleus. Since this is not the case
for all objects we use the optical spectra only for an independent
cross-check of fit-results based on the broad-band colours derived
from our imaging data.
\smallskip

To assess the primary stellar populations of the hosts, we fit stellar
population synthesis model spectra to the multicolour data. For this
we use single age, single metallicity population (SSP) spectra from
the GISSEL96 library (Bruzual \& Charlot 1996, Leitherer et al.\
1996). We chose models with a Salpeter initial mass function and solar
metallicity, ages 0.01--14 Gyr.

These specra were converted to \textsl{BVRiJHKs} colours using ESO
filter curves and fitted to the measured host galaxy colours via a
least-$\chi^2$ fit in two steps: 1) fitting only one SSP, age as free
parameter, 2) fitting two SSPs, ages and mass-ratio of the two
components free.

\section{Results \& Discussion}

The general photometric properties of the sample comply well with
values for inactive galaxies, but of course with a large
object-to-object variation, \textsl{B--V}$=0.76$ (0.78 for an
inactive Sab galaxy), \textsl{V--R}$=0.57$ (0.55 for intermediate type
galaxies), and \textsl{V--K}$=3.2$ (3.2 for intermediate type
galaxies), values taken from Fukugita et al.\ (1995) and Griersmith et
al. (1982).
\begin{figure}[tb]
\hfill
\includegraphics[bb = 107 565 366 771,clip,angle=0,height=7cm]{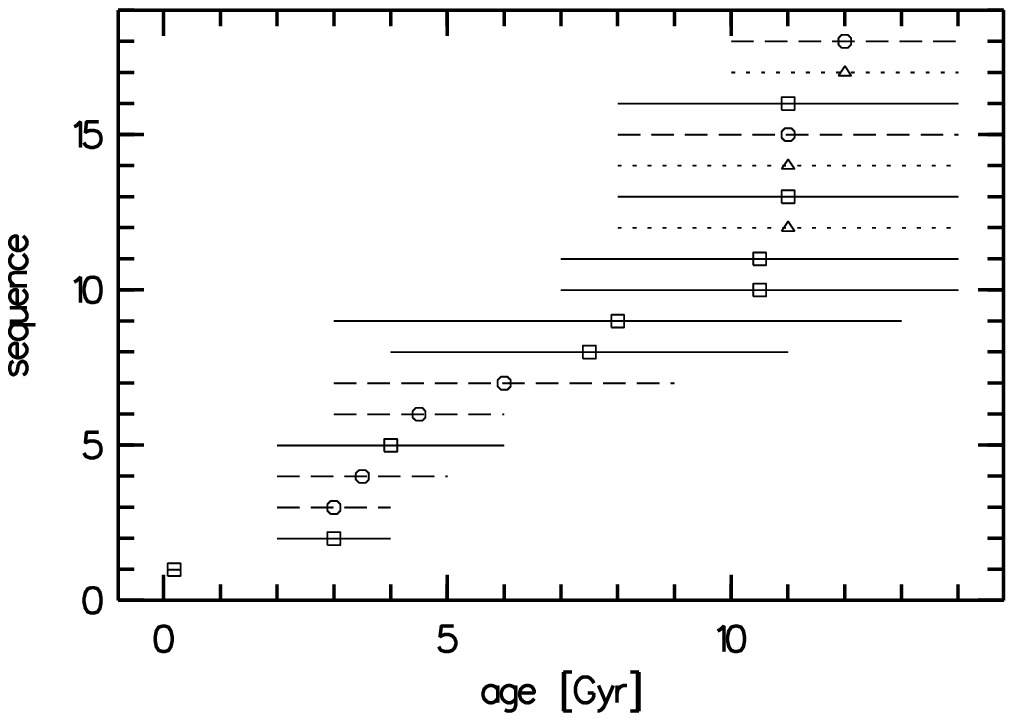}  
\hfill
\caption{\label{1ssp} Fitting one single stellar population
(SSP). Age-sequence of the best fitting SSP for all objects
with range of ages from the fit. Solid lines/squares: morphological
disk; dotted lines/triangles: spheroidal host; dashed lines/circles:
undecided morphological classification.}
\end{figure}

Of the 20 objects we could classify three as spheroidal, ten as disks from
morphological analysis. Seven show signs of at least mild disturbance.
\smallskip

Fitting one or two SSPs is a strong simplification. At least for disks
with a significant amount of onging continuous starformation this will
surely oversimplify the picture. 

Fitting one SSP in principle only compares general optical-to-NIR
colours of host and SSP. Still the ages derived from the fits
(Fig.\ \ref{1ssp}) show a generally good agreement between with ages
expected from the morphological classification. The three classified
spheroidals correspond to old populations of 7--17 Gyr while the
majority of the disks have a clear tendency towards younger SSPs.

\begin{figure}
\hfill
\includegraphics[bb = 74 508 451 774,clip,angle=0,height=7cm]{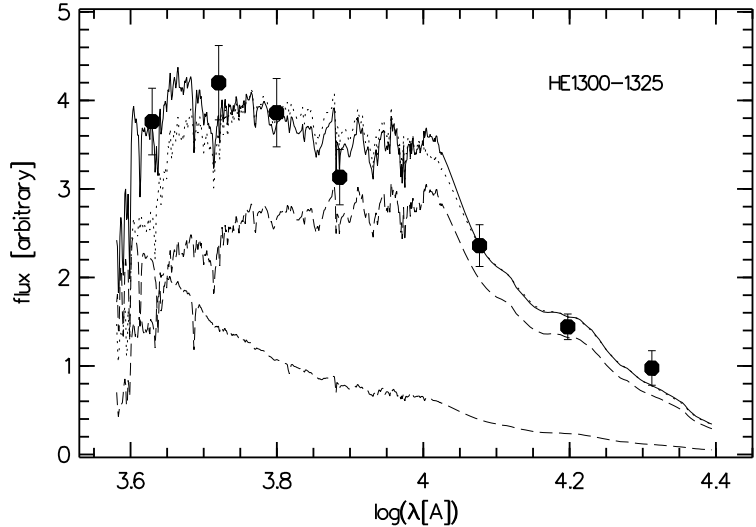}  
\hfill
\caption{\label{he1300} The dotted line respresents one fitted single
stellar population (SSP), the solid line the best combination of two
SSPs, compared to the data points for HE1300-1325. The two lowest
dashed lines are the old and young SSPs contributing to the two-SSP
spectrum. Wavelength scale is log($\lambda$) in {\AA} to better
display the optical part.}
\end{figure}

When fitting two SSPs, we can distinguish contributions from old and
young populations. For most objects though, contributions from a
second population did not improve the fit by a great amount. If at
all, the involved masses of a young population were small, only for
two objects of the order of $\sim2\,\%$. The resulting spectra for one
of them are shown in Fig.\ \ref{he1300}. We see an excess of blue
light in the data (points) compared to the dotted line (single SSP
fit). Since both objects are morphologically classified as disks,
models with continuous star formation will also be able to explain the
blue component.

For the other objects no major second component was detected, and in
fact all these objects are consistent with only one SSP and a uniform
upper limit for the second, younger component of $\sim 0.5\,\%$ (by
mass).

In total we find no signs for strong starburst activity, neither from
the spectral fitting nor from general sample colours. The results are
in favor of the idea that the parent population of QSO host galaxies
is in fact the general field population of inactive galaxies.
\vspace{5mm}

In the future we will use spectral models representing continuous star
formation for disk-type hosts and in addition combine spectral and
colour information into one fitting criterium to make use of all
information available. In order to do this, the contribution from the
current main source of uncertainty, nucleus-galaxy separation, has
to be reduced: we are currently improving our software for spectral
separation to reach a confidence level comparable to that of the
photometric separation.

\end{document}